\documentclass{elsart}
\usepackage{amsfonts}
\usepackage{amssymb}
\usepackage{amsmath}
\usepackage[all]{xy}
\begin{document}

\begin{frontmatter}

\title{On condensation of charged scalars in D=3 dimensions}

\author{Patricio Gaete\thanksref{cile}}
\thanks[cile]{e-mail address: patricio.gaete@usm.cl}
\address{Departamento de F\'{\i}sica and Centro
Cient\'{\i}fico-Tecnol\'{o}gico de Valpara\'{\i}so, Universidad
T\'ecnica Federico Santa Mar\'{\i}a, Valpara\'{\i}so, Chile}
\author{Jos\'{e} A. Hela\"{y}el-Neto\thanksref{cbpf}}
\thanks[cbpf]{e-mail address: helayel@cbpf.br }
\address{Centro Brasileiro de Pesquisas
F\'{\i}sicas, Rua Xavier Sigaud, 150, Urca, 22290-180, Rio de
Janeiro, Brazil}

\begin{abstract}
By using the gauge-invariant, but path-dependent, variables
formalism, we study the impact of condensates on physical
observables for a three-dimensional Higgs-like model. As a result,
for the case of a physical mass term like $m_H^2 \phi ^ *  \phi$, we
recover a screening potential. Interestingly enough, in the case of
a "wrong-sign''   mass term $- m_H^2 \phi ^ *  \phi$, unexpected
features are found. It is shown that the interaction energy is the
sum of an effective-Bessel and a linear potential, leading to the
confinement of static charges. However, when a Chern-Simons term is
included, the surprising result is that the theory describes an
exactly screening phase.
\end{abstract}
\end{frontmatter}

\section{Introduction}
As well-known, a full understanding of the $QCD$ vacuum structure
and color confinement mechanism from first principles remain still
elusive. However, phenomenological models still represent a key tool
for understanding different non-perturbative $QCD$ effects.
Therefore, much about the physics of confinement may be learned from
such models. In this connection it becomes of interest, in
particular, to recall that many approaches to the problem of
confinement rely on the phenomenon of condensation. For example, in
the illustrative  scenario of dual superconductivity \cite{ Nambu,«t
Hooft, Mandelstam}, where it is conjectured that the QCD vacuum
behaves as a dual-type II superconductor. In fact, in this case,
because of the condensation of magnetic monopoles, the
chromo-electric field acting between $q\overline q$ pair is squeezed
into strings, and the nonvanishing string tension
represents the proportionality constant in the linear potential. \\
\\
On the other hand, considerable attention has been paid recently
\cite{Gabadadze1,Gabadadze2, Gabadadze3, Gabadadze4} to condensation
of charged scalars and its physical consequences. The interest in
studying these systems is mainly due to the possibility of
describing condensed helium-4 nuclei in an electron background in
white dwarf cores. More precisely, a Lorentz-violating Higgs-like
effective Lagrangian has been proposed, where a nonzero vacuum
expectation value for the fermion field, which permits to realize
the condensation of the helium-4, plays an essential role in this
development. Accordingly, the condensate characterizes the new
vacuum of the theory with striking consequences over the different
phases of the pure gauge sector of the proposed model. In this
context, in a previous paper \cite{GaeteSpa09}, the impact of
condensates on physical observables in terms of the gauge-invariant
but path-dependent variables formalism has been explored.
Specifically, we have computed the static potential between test
charges in a condensate of scalars and fermions. As a result, in the
case of a "right-sign" mass term $m_H^2 \phi ^ *  \phi$, we have
recovered the screening potential. Interestingly enough, in the case
of a "wrong-sign''   mass term $- m_H^2 \phi ^ *  \phi$, unexpected
features were found. It was observed that the interaction energy is
the sum of an effective-Yukawa and a linear potential, leading to
the confinement of static charges. It is worthwhile mentioning at
this point that the above static profile is analogous to that
encountered in both Abelian and non-Abelian models. For example, in
connection to antisymmetric tensor fields that result from the
condensation of topological defects as a consequence of the
Julia-Toulouse mechanism \cite{GaeteWotzasek04}, in a gauge theory
with a pseudoscalar coupling in the presence of a constant magnetic
strength expectation value \cite{GaeteGuen05}, and in a gauge theory
which includes the mixing between the familiar photon $U(1)_{QED}$
and a second massive gauge field living in the so-called
hidden-sector $U(1)_h$ \cite{GaeteSchmidt08}. Also, in the case of
gluodynamics in curved space-time \cite{GaeteSpallucci08PRD}, and of
a non-Abelian gauge theory with a mixture of pseudoscalar and scalar
couplings, where a constant chromo-electric, or chromo-magnetic,
strength expectation value is present \cite{GaeteSpallucci08JPA}. In
this way, we have provided a new correspondence among diverse
effective theories. This work is devoted to study the stability of
the above scenario for the three-dimensional case. Of special
interest will be to check if a linearly increasing gauge potential
is still present whenever we go over into three dimensions. As well
as, we shall examine the effect of a Chern-Simons term, in the
above scenario, on a physical observable.\\
\\
It is worth recalling at this point that three-dimensional theories
are interesting because of its connection to the high-temperature
limit of four-dimensional theories \cite{Appelquist,Jackiw,Lee,Das}, as
well as, for their applications to condensed matter physics
\cite{Stone}. Thus, as already mentioned, the main purpose here is
to examine the effects of the space-time dimensionality on a
physical observable for the three-dimensional case. To do this, we
will work out the static potential for the model under consideration
by using the gauge-invariant but path-dependent variables formalism
along the lines of Ref. \cite{GaeteSpa09}. As we will see, there are
two generic features that are common in the four-dimensional case
and its lower extension studied here. First, the existence of a
linear potential, leading to the confinement of static charges.
However, when a Chern-Simons term is included, the surprising result
is that the theory describes an exactly screening phase. The second
point is related to the correspondence among diverse effective
theories. In fact, in the case of a "wrong-sign''   mass term $-
m_H^2 \phi ^ *  \phi$, we obtain that the interaction energy is the
sum of an effective-Bessel and a linear potential. Incidentally, the
above static potential profile is analogous to that encountered in:
a Lorentz-and CPT- violating Maxwell-Chern-Simons model
\cite{Helayel}, a Maxwell-like three-dimensional model induced by
the condensation of topological defects driven by quantum
fluctuations \cite{GaeteWot05}, a Lorentz invariant violating
electromagnetism arising from a Julia-Toulouse mechanism
\cite{GaeteWot07}, and three-dimensional gluodynamics in curved
space-time \cite{GaeteHel08}.

Before going ahead, it is appropriate to observe here that a Abelian
gauge theory possessing a confining phase may sound strange. In this
context, it may be recalled that the existence of a phase structure
for the continuum Abelian $U(1)$ gauge theory was obtained by
including the effects due to the compactness of the $U(1)$ group, which
dramatically changes the infrared properties of the model
\cite{Polyakov}. These results, first found in \cite{Polyakov}, have
been ever since rederived by many different techniques
\cite{Ezawa,Orland,Kondo} where the key ingredient is the
contribution of self-dual topological excitations to the partition
function of the theory. However, our analysis renders manifest
that the mechanism of confinement in our model is not condensation
of topological excitations, rather the scalars. This is what makes
the current work different from earlier (above mentioned) proposals
of confinement in Abelian gauge theories.

\section{Three-dimensional Higgs-like model}

As already stated, our principal purpose is to calculate the
interaction energy between static point-like sources for a
Lorentz-violating Higgs-like effective model. To this end, we shall
compute the expectation value of the energy operator $H$ in the
physical state $|\Phi\rangle$, which we will denote by ${\langle
H\rangle}_\Phi$. We begin by summarizing very quickly the recently
proposed Higgs-like model \cite{Gabadadze4,GaeteSpa09}, which
describes a condensed of charged scalars in a neutralizing
background of fermions. This would not only provide the theoretical
setup for our subsequent work, but also fix the notation. The
starting point is the three-dimensional space-time Lagrangian:
\begin{equation}
\mathcal{L} =  - \frac{1}{4}F_{\mu \nu }^2  + \left| {D_\mu  \phi } \right|^2  -
m_H^2 \phi ^ *  \phi + \overline \psi  \left( {i\gamma ^\mu  D_\mu   - M }
\right)\psi ,  \label{ThreeCond05}
\end{equation}
where $\phi$ is a charged massive scalar field, $A_\mu$ is a
$U\left(\,1\,\right)$ gauge potential, and  $\psi$ is an ``heavy''
fermion. The covariant derivative is defined as: $D_\mu \equiv
\partial_\mu +i e A_\mu$. Let us also mention here that $m_H^2>0$ is
a ``right sign'' mass term  and we have not included any
self-interaction for the scalar field. Following our earlier
procedure \cite{GaeteSpa09}, we shall now consider that the fermions
are so heavy that they cannot be excited in the low energy regime we
are studying. In such a case, the Dirac kinetic term can be
neglected and the whole fermion sector of the model reduces  to a
constant background density  $J^0$ coupled to $A_\mu$, that is,
$\overline \psi \gamma ^\mu \psi\longrightarrow -\delta^\mu_0\,
J^0$. This allows us to write the Lagrangian (\ref{ThreeCond05}) as
\begin{equation}
\mathcal{ L} = - \frac{1}{4}F_{\mu \nu }^2  + \left| {D_\mu  \phi } \right|^2
- m_H^2 \phi ^ *  \phi -e J^0  \delta^\mu_0\, A_\mu.  \label{ThreeCond10}
\end{equation}

Once this is done, the field equations obtained by varying
(\ref{ThreeCond10}) with respect to $A_\mu$ and $\phi^\ast$ follow
closely that of reference \cite{GaeteSpa09}:
\begin{equation}
\partial_\mu F^{\mu\nu} +2e^2 A^\nu |\phi|^2 = e(J^\nu_s + J^0  \delta^\nu_0),
\label{ThreeCond15}
\end{equation}
\begin{equation}
\left(\, \partial_\mu \partial^\mu - e^2 A_\mu A^\mu + m^2_H \,\right)\phi =0,
\label{ThreeCond20}
\end{equation}

where $J^\nu_s\equiv i(\phi^\ast\partial^\nu \phi -\phi\partial^\nu
\phi^\ast)$. In this way, the  ground state of the system is
described by the classical solution:
\begin{equation}
\overline{\psi}_0 \gamma ^\mu \psi_0= -\delta^\mu_0\ J^0,\label{ThreeCond25}
\end{equation}
\begin{equation}
\phi_0=\sqrt{\frac{J^0}{2m_H}},\label{ThreeCond30}
\end{equation}
\begin{equation}
A_{0}^{\mu }=\frac{m_H}{e} \,  \delta_{0}^{\mu}.\label{ThreeCond35}
\end{equation}

Once there is a non-vanishing background value for the scalar field,
we choose to work in the unitary gauge, so that the phase of the
$\phi$ - field can be gauged away. Next to this choice, we split the
fields $\phi$ (now, $\phi =\phi^*$)   and $A_\mu$ as the sum of a
classical background around which there appear quantum fluctuations
as it follows below:
\begin{equation}
\phi =\phi^*= \phi_0 +\frac{1}{\sqrt{2}}\eta\left(\, x\,\right),
\label{ThreeCond40}
\end{equation}
\begin{equation}
A_\mu =\frac{m_H}{e} \,  \delta^0_\mu + b_\mu\left(\, x\,\right),
\label{ ThreeCond45}
\end{equation}
the corresponding Lagrangian density, up to quadratic terms in the
fluctuations, is given by
\begin{equation}
\mathcal{ L} =  - \frac{1}{4}f_{\mu \nu }^2  + \frac{1}{2}\left(
{\partial _\mu \eta } \right)^2 + \frac{1}{2}m_\gamma ^2 b_\mu ^2  +
2m_H m_\gamma b_0\,  \eta\ . \label{ThreeCond50}
\end{equation}
where $f_{\mu \nu } \equiv \partial _\mu  b_\nu   - \partial _\nu
b_\mu$, and $m^2_\gamma\equiv 2e^2\phi_0^2$. Following our earlier
procedure \cite{GaeteSpa09}, integrating out the $\eta$ field
induces an effective theory for the  $b_\mu$ field. This leads us to
the following effective Lagrangian density:
\begin{equation}
\mathcal{ L}_{eff} =  - \frac{1}{4}f_{\mu \nu }^2  +
\frac{1}{2}m_\gamma ^2 b_\mu ^2 +2m_H^2 m_\gamma^2\, b_0\,
\frac{1}{\Delta }\,b_0, \label{ThreeCond55}
\end{equation}
where $\Delta=\partial_\mu\partial^\mu$. As a consequence, the
Lagrangian (\ref{ThreeCond05}) becomes a Maxwell-Proca-like theory
with a manifestly Lorentz violating term. This effective theory
provide us with a suitable starting point to study the interaction
energy. However, before proceeding with the determination of the
energy, it is necessary to restore the gauge invariance in
(\ref{ThreeCond55}). For this purpose, we note that the Lagrangian
(\ref{ThreeCond55}) may be rewritten as
\begin{equation}
{\cal L} =  - \frac{1}{4}f_{\mu \nu }^2  + \frac{1}{2}b_\mu  m^2
b^\mu - \frac{1}{2}b_i \frac{{\left( {2m_H m_\gamma  } \right)^2
}}{\Delta }b^i , \label{ThreeCond60}
\end{equation}
where $m^2  \equiv m_\gamma ^2 \left( {1 + \frac{{4m_H^2 }}{\Delta
}} \right)$. With this in hand, the canonical momenta $\Pi^\mu$ are
found to be $\Pi ^0=0$ and $\Pi ^i =  - f^{0i}$. The canonical
Hamiltonian is now obtained in the usual way
\begin{equation}
H = \int {d^2 x} \left\{ { - b_0 \left( {\partial _i \Pi ^i  +
\frac{{m^2 }}{2}b^0 } \right) - \frac{1}{2}\Pi _i \Pi ^i  +
\frac{1}{4}f_{ij} f^{ij}  - \frac{1}{2}b_i \left( {m^2  -
\frac{{\left( {2m_H m_\gamma  } \right)}}{\Delta }} \right)b^i }
\right\}. \label{ThreeCond65}
\end{equation}\\
Time conservation of the primary constraint ($\Pi ^0=0$) yields a
secondary constraint $\Gamma \left( x \right) \equiv \partial _i \Pi
^i  + m^2 b^0  = 0$. Notice that the nonvanishing bracket $\left\{
{\Pi ^0 ,\partial _i \Pi ^i  + m^2 b^0 } \right\}$ shows that the
above pair of constraints are second class constraints, as expected
for a theory with an explicit mass term which breaks the gauge
invariance. To convert the second class system into first class we
enlarge the original phase space by introducing a canonical pair of
fields $\theta$ and $\Pi _\theta$ \cite{GaeteSpa09}. It follows,
therefore, that a new set of first class constraints can be defined
in this extended space:
\begin{equation}
\Lambda _1  \equiv \Pi _0  + m^2 \theta, \label{ThreeCond70a}
\end{equation}
and
\begin{equation}
\Lambda _2  \equiv \Gamma  + \Pi _\theta. \label{ThreeCond70b}
\end{equation}
In this way the gauge symmetry of the theory under consideration has
been restored. Then, the new effective Lagrangian, after integrating
out the $\theta$ field, becomes
\begin{equation}
{\cal L}_{eff} = - \frac{1}{4}f_{\mu \nu } \left[ {1 +
\frac{{m_\gamma ^2 }} {\Delta }\left( {1 + \frac{{4m_H^2 }}{\Delta
}} \right)} \right]f^{\mu \nu }. \label{ThreeCond80}
\end{equation}
Again, as was explained in \cite{GaeteSpa09}, we observe that to get
the above theory we have ignored the last term in
(\ref{ThreeCond60}) because it add nothing to the static potential
calculation, as we will show it below. In other words, the new
effective action (\ref{ThreeCond80})  provide us with a suitable
starting point to study the interaction energy without loss of
physical content.

We now turn our attention to the calculation of the interaction
energy. In order to obtain the corresponding Hamiltonian, the
canonical quantization of this theory from the Hamiltonian analysis
point of view is straightforward and follows closely that of
reference \cite{GaeteSpa09}. The canonical momenta read $\Pi ^\mu
=  - \left[ {1 + \frac{{m_\gamma ^2 }}{\Delta }\left( {1 +
\frac{{4m_H^2 }}{\Delta }} \right)} \right]f^{0\mu }$, and one
immediately identifies the usual primary constraint $\Pi ^0  = 0$
and $ \Pi ^i  =  - \left[ {1 + \frac{{m_\gamma ^2 }}{\Delta }\left(
{1 + \frac{{4m_H^2 }}{\Delta }} \right)} \right]f^{0i}$. The
canonical Hamiltonian is thus
\begin{equation}
H_C  = \int {d^2 x} \left\{ { - b_0 \partial _i \Pi ^i  -
\frac{1}{2}\Pi _i \left[ {1 + \frac{{m_\gamma ^2 }}{\Delta }\left(
{1 + \frac{{4m_H^2 }}{\Delta }} \right)} \right]^{ - 1} \Pi ^i  +
\frac{1}{4}f_{ij} f^{ij} } \right\}. \label{ThreeCond85}
\end{equation}
The consistency condition $\dot \Pi _0  = 0$ leads to the usual Gauss
constraint $\Gamma_1 \left( x \right) \equiv \partial _i \Pi ^i=0$.
The extended Hamiltonian that generates translations in time then reads
$H = H_C + \int {d^2}x\left( {c_0 \left( x \right)\Pi _0 \left( x \right)
 + c_1 \left(x\right)\Gamma _1 \left( x \right)} \right)$, where $c_0 \left(
x\right)$ and $c_1 \left( x \right)$ are the Lagrange
multipliers. Since $ \Pi^0 = 0$ for all time and $\dot{b}_0 \left( x \right)=
\left[ {b_0 \left( x \right),H} \right] = c_0 \left( x \right)$,
which is completely arbitrary, we discard $b^0$ and $\Pi^0$ because they adding
nothing to the description of the system. Then, the Hamiltonian takes the form
\begin{equation}
H = \int {d^2 x} \left\{ {  c(x) \partial _i \Pi ^i  - \frac{1}{2}\Pi _i \left[ {1 +
\frac{{m_\gamma ^2 }}{\Delta }\left( {1 + \frac{{4m_H^2 }}{\Delta }} \right)}
\right]^{ - 1} \Pi ^i  + \frac{1}{4}f_{ij} f^{ij} } \right\}, \label{ThreeCond90}
\end{equation}
where $c(x) = c_1 (x) - b_0 (x)$. Evidently, the presence of the
arbitrary quantity $c(x)$ is undesirable since we have no way to
giving it a meaning in a quantum theory. As is well known, the
solution to this problem is to introduce a gauge condition such that
the full set of constraints become second class. A particularly
convenient choice is found to be
\begin{equation}
\Gamma _2 \left( x \right) \equiv \int\limits_{C_{\xi x} } {dz^\nu }
b_\nu \left( z \right) \equiv \int\limits_0^1 {d\lambda x^i } b_i
\left( {\lambda x} \right) = 0,     \label{ThreeCond95}
\end{equation}
where  $\lambda$ $(0\leq \lambda\leq1)$ is the parameter describing
the spacelike straight path $ x^i = \xi ^i  + \lambda \left( {x -
\xi } \right)^i $, and $ \xi $ is a fixed point (reference point).
There is no essential loss of generality if we restrict our
considerations to $ \xi ^i=0 $. The choice (\ref{ThreeCond95}) leads to
the Poincar\'e gauge \cite{Gaete99}. As a consequence, we can
now write down the only nonvanishing Dirac bracket for the canonical
variables
\begin{equation}
\left\{ {b_i \left( x \right),\Pi ^j \left( y \right)} \right\}^ *
=\delta{ _i^j} \delta ^{\left( 2 \right)} \left( {x - y} \right) -
\partial _i^x \int\limits_0^1 {d\lambda x^j } \delta ^{\left( 2
\right)} \left( {\lambda x - y} \right). \label{ThreeCond100}
\end{equation}

We are now ready to find  the interaction energy between point-like
sources for the model under consideration. As we have already indicated, we will
calculate the expectation value of the energy operator $H$ in the physical
state $|\Phi\rangle$. In this context, we recall that the physical state
$|\Phi\rangle$ can be written as
\begin{equation}
\left| \Phi  \right\rangle  \equiv \left| {\overline \Psi  \left(
\bf y \right)\Psi \left( {\bf 0} \right)} \right\rangle
= \overline \psi \left( \bf y \right)\exp \left(
{iq\int\limits_{{\bf 0}}^{\bf y} {dz^i } b_i \left( z \right)}
\right)\psi \left({\bf 0} \right)\left| 0 \right\rangle,
\label{ThreeCond105}
\end{equation}
where $\left| 0 \right\rangle$ is the physical vacuum state. The line integral
is along a spacelike path starting at $\bf 0$ and ending at $\bf y$, on a fixed
time slice.

Next, taking into account the above Hamiltonian analysis, we then obtain
\begin{equation}
\left\langle H \right\rangle _\Phi   = \left\langle H \right\rangle
_0 + \left\langle H \right\rangle _\Phi ^{\left( 1 \right)},
\label{ThreeCond110}
\end{equation}
where, in this static case, $\Delta = -\nabla^{2}$. At the same time,
$\left\langle H \right\rangle _0  = \left\langle 0 \right|H\left| 0 \right\rangle$,
and the $\left\langle H \right\rangle _\Phi ^{\left( 1 \right)}$ term is given by
\begin{equation}
\left\langle H \right\rangle _\Phi ^{\left( 1 \right)}  =
\left\langle \Phi \right|\int {d^2 x} \left\{ { - \frac{1}{2}\Pi _i
\left[ {1 - \frac{{m_\gamma ^2 }}{{\nabla ^2 }}\left( {1 -
\frac{{4m_H^2 }}{{\nabla ^2 }}} \right)} \right]^{ - 1}\Pi ^ i  +
\frac{1}{4} f_{ij}f^{ij} } \right\}\left| \Phi  \right\rangle .
\label{ThreeCond115}
\end{equation}
It should be noted that the above expression may be rewritten as
\begin{eqnarray}
\left\langle H \right\rangle _\Phi ^{\left( 1 \right)}  &=&  -
\frac{1}{2} \frac{{4M^4 }}{{\left( {M_2^2  - M_1^2 } \right)}}\int
{d^2 } x\left\langle \Phi  \right|\Pi _i \left\{ {\alpha
\frac{{\nabla ^2 }}{{\left( {\nabla ^2  - M_1^2 } \right)}} - \beta
\frac{{\nabla ^2 }}{{\left( {\nabla ^2  - M_2^2 } \right)}}}
\right\}\Pi ^i \left| \Phi  \right\rangle + \nonumber\\
&+& \frac{1}{4}\int {d^2 x} \left\langle \Phi  \right|f_{ij} f^{ij}
\left| \Phi  \right\rangle, \label{ThreeCond120}
\end{eqnarray}
with $\alpha  = \frac{1}{{\left( {M_1^2  - m_\gamma ^2 } \right)}}$ and $
\beta  = \frac{1}{{\left( {M_2^2  - m_\gamma ^2 } \right)}}$.  While
$M_1^2  = \frac{1}{2}\left( {m_\gamma ^2  + \sqrt {m_\gamma ^4  - 16M^4 } }
 \right)$,\\
$M_2^2  = \frac{1}{2}\left( {m_\gamma ^2  - \sqrt {m_\gamma ^4  - 16M^4 } }
 \right)$ and $M \equiv \sqrt {m_\gamma  m_H } $. One immediately sees that this
expression is similar to that encountered in the three space
dimensions case \cite{GaeteSpa09}. It follows, therefore, that in
$(2+1)$ dimensions, the potential for two opposite charges located
at ${\bf 0}$ and $\bf y$ takes the form
\begin{equation}
V =  - \frac{{q^2 }}{{2\pi }}\frac{{4M^4 }}{{\sqrt {m_\gamma ^4  -
16M^4 } }}\left[ {\frac{1}{{M_2^2 }}K_0 \left( {M_1 L} \right) +
\frac{1}{{M_1^2 }}K_0 (M_2 L)} \right], \label{ThreeCond125}
\end{equation}
where $K_0$ is a modified Bessel function, and $|{\bf y}|\equiv L$.

Before we proceed further, we wish to illustrate an alternative
derivation of the result (\ref{ThreeCond125}), which exhibits
certain distinctive features of our methodology. To begin with, let
us recall that the potential can be obtained from \cite{Pato}:
\begin{equation}
V \equiv q\left( {{\cal A}_0 \left( {\bf 0} \right) - {\cal A}_0
\left( {\bf y} \right)} \right), \label{ThreeCond130}
\end{equation}
where the physical scalar potential is given by
\begin{equation}
{\cal A}_0 \left( {x^0 ,{\bf x}} \right) = \int_0^1 {d\lambda } x^i
E_i \left( {\lambda {\bf x}} \right), \label{ThreeCond135}
\end{equation}
with $i=1,2$. It is worth noting here that this follows from the
vector gauge-invariant field expression \cite{Pato2}
\begin{equation}
{\cal A}_\mu  \left( x \right) \equiv A_\mu  \left( x \right) +
\partial _\mu  \left( { - \int_\xi ^x {dz^\mu  } A_\mu  \left( z
\right)} \right), \label{ThreeCond140}
\end{equation}
where, as in Eq.(\ref{ThreeCond95}), the line integral is along a
spacelike path from the point $\xi$ to $x$, on a fixed slice time.
The gauge-invariant variables (\ref{ThreeCond140}) commute with the sole
first constraint (Gauss' law), confirming that these fields are
physical variables \cite{Dirac2}. Note that Gauss' law for the
present theory reads $\partial _i \Pi ^i  = J^0$, where we have
included the external current $J^0$ to represent the presence of two
opposite charges. For $J^0 \left( {t,{\bf x}} \right) = q\delta
^{\left( 2 \right)} \left( {\bf x} \right)$ the electric field is
given by
\begin{equation}
E^i  = q\frac{{4M^4 }}{{\left( {M_2^2  - M_1^2 } \right)}}\left\{
{\frac{1} {{\left( {M_1^2  - m_\gamma ^2 } \right)}}\partial ^i
G^{\left( 1 \right)} \left( \bf x \right) - \frac{1}{{\left( {M_2^2
- m_\gamma ^2 } \right)}}\partial ^i G^{\left( 2 \right)} \left( \bf
x \right)} \right\}, \label{ThreeCond145}
\end{equation}
where $G^{\left( 1 \right)} \left( \bf x \right) = \frac{1}{{2\pi
}}K_0 \left( {M_1 \left| {\bf x} \right|} \right)$ and $G^{\left( 2
\right)} \left( \bf x \right) = \frac{1}{{2\pi }}K_0 \left( {M_2
\left| {\bf x} \right|} \right)$ are the Green functions for the
Proca operator in two space dimensions. Using this, the physical
scalar potential, Eq.(\ref{ThreeCond135}), reduces to
\begin{equation}
{\cal A}_0 \left( {t,{\bf x}} \right) = q\frac{{4M^4 }}{{\left(
{M_2^2  - M_1^2 } \right)}}\left[ {\frac{1}{{\left( {M_1^2  -
m_\gamma ^2 } \right)}}G^{\left( 1 \right)} \left( \bf x \right) -
\frac{1}{{\left( {M_2^2  - m_\gamma ^2 } \right)}}G^{\left( 2
\right)} \left( \bf x \right)} \right], \label{ThreeCond150}
\end{equation}
after substraction of self-energy terms. With this then, we now see
that the potential for a pair of point-like opposite charges q
located at ${\bf 0}$ and ${\bf L}$ becomes
\begin{equation}
V =  - \frac{{q^2 }}{{2\pi }}\frac{{4M^4 }}{{\sqrt {m_\gamma ^4  -
16M^4 } }}\left[ {\frac{1}{{M_2^2 }}K_0 \left( {M_1 L} \right) +
\frac{1}{{M_1^2 }}K_0 (M_2 L)} \right], \label{ThreeCond155}
\end{equation}
where $\left| {\bf L} \right| \equiv L$. It must be clear from this
discussion that a correct identification of physical degrees of
freedom is a key feature for understanding the physics hidden in
gauge theories.

Within this framework, we now want to extend what we have done when
a $m_H^2 \phi ^ *  \phi$ term and a quartic self-interaction
potential is considered in expression (\ref{ThreeCond05}), namely,
\begin{equation}
\mathcal{L} =  - \frac{1}{4}F_{\mu \nu }^2  + |D_\mu  \phi |^2 +
m_H^2 \phi ^ *  \phi  - \frac{\lambda }{6}\left( {\phi ^ *  \phi }
\right)^2  - eJ^0 \delta _\mu ^0 A_\mu .  \label{ThreeCond160}
\end{equation}
As before, the last term arises from the condensation mechanism in a
neutralizing background of fermions. Following the same steps that
lead to (\ref{ThreeCond80}) we arrive at the following effective
Lagrangian density:
\begin{equation}
{\mathcal L}_{eff}  =  - \frac{1}{4}f_{\mu \nu } \left[ {1 +
\frac{{m_\gamma ^2 }}{\Delta }\left( {1 + \frac{{4\mu _s^2
}}{{\left( {\Delta  + 2m_H^2 } \right)}}} \right)} \right]f^{\mu \nu
}. \label{ThreeCond165}
\end{equation}
In the same way as was done in the previous case, one finds
\begin{eqnarray}
\left\langle H \right\rangle _\Phi   &=& C_1 \left\langle \Phi  \right|
- \frac{1}{2}\int {d^2 x} \Pi _i \left\{ {\frac{{\nabla ^2 }}
{{\left( {\nabla ^2  - M_2^2 } \right)}} - \eta ^2 \frac{{\nabla ^2 }}
{{\left( {\nabla ^2  - M_1^2 } \right)}}} \right\}\Pi ^i \left| \Phi
\right\rangle  \nonumber \\
&+& C_2 \left\langle \Phi  \right|\frac{1}{2}\int {d^2 x} \Pi _i
\left\{ {\frac{1}{{\left( {\nabla ^2  - M_2^2 } \right)}} - \eta ^2
\frac{1}{{\left( {\nabla ^2  - M_1^2 } \right)}}} \right\}\Pi ^i
\left| \Phi  \right\rangle,  \label{ThreeCond170}
\end{eqnarray}
where $C_1  \equiv \frac{{2M^4 }}{{\left( {2M^4  - m_\gamma ^2 } \right)}}$,
$C_2  \equiv \frac{{4m_H^2 M^4 }}{{\left( {2M^4  - m_\gamma ^2 } \right)}}$, and $
\eta ^2  \equiv \frac{{m_\gamma ^2 }}{{2m_H^2 }}$. While $M_1^2  = m_\gamma ^2$,
$M_2^2  = 2m_H^2$, and $M = \sqrt {m_\gamma  m_H }$.

According to our earlier procedure, we find that the potential for
two opposite charges located at ${\bf 0}$ and $\bf y$ takes the form
\begin{equation}
V =  - \frac{{q^2 }}{{2\pi }}C_1 \left\{ {K_0 \left( {M_2 L} \right)
- \frac{{M_1^2 }}{{M_2^2 }}K_0 \left( {M_1 L} \right)} \right\} +
\frac{{q^2 }}{4}\frac{{C_2 }}{{M_2 }}\left\{ {1 - \frac{{M_1 }}{{M_2
}}} \right\}L. \label{ThreeCond175}
\end{equation}
Here, in contrast to the previous case, unexpected features are
found. In fact, we see that the static potential profile displays
the conventional screening part, encoded in the modified Bessel
function, and the linear confining potential.

\section{Three-dimensional Higgs-like model and a Chern-Simons term}

We now pass on to the calculation of the interaction energy between
static pointlike sources for the $(2+1)$-dimensional Higss-like
model with a Chern-Simons term. In other words, in this section we
concentrate on the effect of including the Chern-Simons term in the
confinement and screening nature of the potential. With this in
mind, we start by writing:
\begin{equation}
{\mathcal L} =  - \frac{1}{4}F_{\mu \nu }^2  +
\frac{s}{2}\varepsilon ^{\nu \kappa \lambda } A_\nu  \partial
_\kappa  A_\lambda   + |D_\mu  \phi |^2  - m_H^2 \phi \phi ^ *   -
eJ^0 \delta _0^\mu  A_\mu . \label{ThreeCond180}
\end{equation}
Proceeding as in the previous subsection, the effective Lagrangian
is given by
\begin{equation}
{\mathcal L}_{eff} =  - \frac{1}{4}f_{\mu \nu } \left[ {1 +
\frac{{m_\gamma ^2 }} {\Delta }\left( {1 + \frac{{4m_H^2 }}{\Delta
}} \right)} \right]f^{\mu \nu } + \frac{s}{2}\varepsilon ^{\mu \nu
\lambda } b_\mu  \partial _\nu  b_\lambda. \label{ThreeCond185}
\end{equation}

The effective Lagrangian expressed by (\ref{ThreeCond185}) describes
the effective dynamics of the quantum $b_\mu$-field. Since we are
interested in pursuing an investigation of the potential which comes
from the  $b_\mu$-field exchange, we can say that we are actually
restricting our analysis to the low-frequency regime of ${\mathcal
L}_{eff}$. In this region, it is legitimate to drop the  $f_{\mu \nu
}f^{\mu \nu }$-term respect to the other terms, the reason being
that this term is quadratic in the frequencies and, therefore, the
terms  $m_\gamma ^2$  and $s$ dominate. The space-time
dependence of $b_\mu$ and, hence, its dynamics, is accounted for
in the $f_{\mu \nu}^2$ and in the Chern-Simons terms. Considering the regime of low-frequencies, it is true that they are both much smaller than the term in $m^2$. However, disregarding them simultaneously would
lead us to a completely different regime, where only constant field configurations would be considered. To ensure that contributions from
non-constant configurations are also taken into account, we have to keep at the least the Chern-Simons term, since it is linear in the frequency whereas the Maxwell-term is quadratic. So, our claim is that the s-term
is the one that survives in the low-frequency regime, and this guarantees that non-constant field configurations are not thrown away. Therefore, keeping in mind that we are bound to the low-frequency regime, we can express ${\mathcal L}_{eff}$ as follows:
\begin{equation}
{\mathcal L}_{eff} =  - \frac{1}{4}f_{\mu \nu } \left[ {
\frac{{m_\gamma ^2 }} {\Delta }\left( {1 + \frac{{4m_H^2 }}{\Delta
}} \right)} \right]f^{\mu \nu } + \frac{s}{2}\varepsilon ^{\mu \nu
\lambda } b_\mu  \partial _\nu  b_\lambda. \label{ThreeCond185b}
\end{equation}

It is now once again straightforward to apply the gauge-invariant
formalism discussed in the preceding section. For this purpose, we
start by observing that the canonical momenta read $ \Pi ^\mu   =
\left[ {\frac{{m_\gamma ^2 }}{\Delta }\left( {1 + \frac{{4m_H^2
}}{\Delta }} \right)} \right]f^{\mu 0}  + \frac{s}{2}\varepsilon
^{0\mu \nu } b_\nu$. As we can see there is one primary constraint
$\Pi ^0=0$, and $\Pi ^i   = \left[ {\frac{{m_\gamma ^2 }}{\Delta
}\left( {1 + \frac{{4m_H^2 }}{\Delta }} \right)} \right]f^{i 0}  +
\frac{s}{2}\varepsilon ^{i j} b_j$. The canonical Hamiltonian for
this system, in terms of $B = \varepsilon _{ij} \partial ^i b^j$ and
$ E^i  = \left[ {\frac{{m_\gamma ^2 }}{\Delta }\left( {1 +
4\frac{{m_H^2 }}{\Delta }} \right)}\right]^{ - 1} (\Pi ^i  -
\frac{s}{2}\varepsilon ^{ij} b_j)$, is in this case
\begin{eqnarray}
H_C  &=& \int {d^2 x} \left\{ { - b^0 \left( {\partial _i \Pi ^i  +
 \frac{s}{2}B} \right) + \frac{1}{2}E^i \left[ {\frac{{m_\gamma ^2 }}
{\Delta }\left( {1 + \frac{{4m_H^2 }}{\Delta }} \right)} \right]E^i } \right\}
\nonumber\\
&+& \int {d^2 x} \left\{ {\frac{1}{2}B\left[ {\frac{{m_\gamma ^2
}} {\Delta }\left( {1 + \frac{{4m_H^2 }}{\Delta }} \right)}
\right]B} \right\}. \label{ThreeCond190}
\end{eqnarray}
The conservation in time of the primary constraint $\Pi^0$
leads to the secondary constraint $\Gamma _1 \left( x \right) \equiv
\partial _i \Pi ^i  + \frac{s}{2}B = 0$. The above constraints are the
first-class constraints of the theory since no more constraints are
generated by the time preservation of the secondary constraints.
Once again, the corresponding total (first-class) Hamiltonian that
generates the time evolution of the dynamical variables reads $H =
H_C + \int {d^2}x\left( {c_0 \left( x \right)\Pi _0 \left( x \right)
+ c_1 \left(x\right)\Gamma _1 \left( x \right)} \right)$, where $c_0
\left(x\right)$ and $c_1 \left( x \right)$ are the Lagrange
multiplier fields to implement the constraints. As before, neither
$b_0(x)$ nor $\Pi _0(x)$ are of interest in describing the system
and may be discarded from the theory. As a result the Hamiltonian
becomes
\begin{eqnarray}
H  &=& \int {d^2 x} \left\{ { c(x) \left( {\partial _i \Pi ^i  +
\frac{s}{2}B} \right) + \frac{1}{2}E^i \left[ {\frac{{m_\gamma ^2 }}
{\Delta }\left( {1 + \frac{{4m_H^2 }}{\Delta }} \right)} \right]E^i }
\right\} \nonumber\\
&+& \int {d^2 x} \left\{ {\frac{1}{2}B\left[ {\frac{{m_\gamma ^2 }}
{\Delta }\left( {1 + \frac{{4m_H^2 }}{\Delta }} \right)} \right]B} \right\},
 \label{ThreeCond195}
\end{eqnarray}
where $c(x) = c_1 (x) - b_0 (x)$. Since our main motivation is to
compute the static potential, we will adopt the same gauge-fixing
condition that was used in the last subsection. Thus, in order to
illustrate the discussion, we now write the Dirac brackets in terms
of the magnetic and electric fields as:
\begin{equation}
\left\{ {E_i \left( x \right),E_j \left( y \right)} \right\}^ *   =
- s\left[ {\frac{{m_\gamma ^2 }}{\Delta }\left( {1 +
\frac{{4m_H^2 }}{\Delta }} \right)} \right]^{ - 2} \varepsilon _{ij}
\delta ^{\left( 2 \right)} \left( {x - y} \right),
\label{ThreeCond200a}
\end{equation}

\begin{equation}
\left\{ {B\left( x \right),B\left( y \right)} \right\}^ *   = 0,
\label{ThreeCond200b}
\end{equation}

\begin{equation}
\left\{ {E_i \left( x \right),B \left( y \right)} \right\}^ *   = -
\left[ {\frac{{m_\gamma ^2 }}{\Delta }\left( {1 + \frac{{4m_H^2
}}{\Delta }} \right)} \right]^{ - 1} \varepsilon
_{ij}\partial_{x}^{j} \delta ^{\left( 2 \right)} \left( {x - y}
\right).  \label{ThreeCond200c}
\end{equation}

One can now easily derive the equations of motion for the electric
and magnetic fields. We find
\begin{equation}
\dot E_i \left( x \right) =  - s\left[ {\frac{{m_\gamma ^2 }}
{\Delta }\left( {1 + \frac{{4m_H^2 }}{\Delta }} \right)} \right]^{ -
1} \varepsilon _{ij} E_j \left( x \right) - \varepsilon _{ij}
\partial ^j B\left( x \right), \label{ThreeCond205}
\end{equation}
\begin{equation}
\dot B\left( x \right) =  - \varepsilon _{ij} \partial ^j E_i \left(
x \right). \label{ThreeCond210}
\end{equation}

In the same way, we write the Gauss law as
\begin{equation}
\left[ {\frac{{m_\gamma ^2 }}{\Delta }\left( {1 + \frac{{4m_H^2
}} {\Delta }} \right)} \right]\partial _i E^i  + sB + J^0  = 0.
\label{ThreeCond215}
\end{equation}
As before, we have included the external current $J^0$ to represent
the presence of two opposite charges. For $J^0 \left( {t,{\bf x}}
\right) = q\delta ^{\left( 2 \right)} \left( {\bf x} \right)$ the
electric field, in the $m_H^2  \gg {\bf k}^2 $ case,  is given by
\begin{equation}
E^i  = \frac{q}{{m_\gamma ^2 \sqrt {1 - 32s^2 {{m_H^2 } \mathord{\left/
 {\vphantom {{m_H^2 } {m_\gamma ^4 }}} \right.
 \kern-\nulldelimiterspace} {m_\gamma ^4 }}} }}\left[ {\alpha
\partial ^i G^{\left( 2 \right)} \left( {\bf x} \right) - \beta
\partial ^i G^{\left( 1 \right)} \left( {\bf x} \right)} \right],
\label{ThreeCond215}
\end{equation}
where $\alpha  = M_2^2  - 4m_H^2 $, $ \beta  = M_1^2  - 4m_H^2 $, $
M_1^2  = \frac{{m_\gamma ^4 }}{{2s^2 }}\left[ {1 + \sqrt {1 - 32s^2
{{m_H^2 } \mathord{\left/ {\vphantom {{m_H^2 } { m_\gamma ^4 }}}
\right.
 \kern-\nulldelimiterspace} {m_\gamma ^4 }}} } \right]$, and $
M_2^2  = \frac{{m_\gamma ^4 }}{{2s^2 }}\left[ {1 - \sqrt {1 - 32s^2
{{m_H^2 } \mathord{\left/ {\vphantom {{m_H^2 } {m_\gamma ^4 }}}
\right.
 \kern-\nulldelimiterspace} {m_\gamma ^4 }}} } \right]$. Again,
$G^{\left( 1 \right)} ({\bf x}) = \frac{1}{{2\pi }}K_0 (M_1 |{\bf
x}|)$, and $G^{\left( 2 \right)} ({\bf x}) = \frac{1}{{2\pi }}K_0
(M_2 |{\bf x}|)$.  Combining Eqs. (\ref{ThreeCond215}) and
(\ref{ThreeCond130}), we can write immediately the potential for a
pair of point-like opposite charges q located at ${\bf 0}$ and ${\bf
L}$, as
\begin{equation}
V =  - \frac{{q^2 }}{{2\pi }}\frac{1}{{m_\gamma ^2 \sqrt {1 - 32s^2
{{m_H^2 } \mathord{\left/
 {\vphantom {{m_H^2 } {m_\gamma ^4 }}} \right.
 \kern-\nulldelimiterspace} {m_\gamma ^4 }}} }}\left[ {\alpha K_0
\left( {M_2 L} \right) - \beta K_0 (M_1 L)} \right],
\label{ThreeCond220}
\end{equation}
where $|{\bf L}| = L$.

Let us consider next the effect of a $m_H^2 \phi ^ *  \phi$ term and
a quartic self-interaction potential in expression
(\ref{ThreeCond180}), that is,
\begin{equation}
{\mathcal L} =  - \frac{1}{4}F_{\mu \nu }^2  +
\frac{s}{2}\varepsilon ^{\nu \kappa \lambda } A_\nu  \partial
_\kappa  A_\lambda   + |D_\mu  \phi |^2  + m_H^2 {\phi}^{\ast}\phi -
\frac{\lambda}{6}({\phi}^{\ast}\phi)^2 - eJ^0 \delta _0^\mu  A_\mu .
\label{ThreeCond225}
\end{equation}

Again, in the same way as was done in the previous case, one finds
\begin{equation}
{\mathcal L}_{eff}  =  - \frac{1}{4}f_{\mu \nu } \left[ {1 +
\frac{{m_\gamma ^2 }}{\Delta }\left( {1 + \frac{{4\mu _s^2
}}{{\left( {\Delta  + 2m_H^2 } \right)}}} \right)} \right]f^{\mu \nu
}  + \frac{s}{2}\varepsilon ^{\mu \nu \lambda } b_\mu  \partial _\nu
b_\lambda. \label{ThreeCond230}
\end{equation}

Again, as discussed in going from Eq. (\ref{ThreeCond185}) to Eq.
(\ref{ThreeCond185b}), we here also work in the regime of low
frequencies, so that it the $f_{\mu \nu }f^{\mu \nu}$-term can be
neglected in comparison with the other terms.

\begin{equation}
{\mathcal L}_{eff}  =  - \frac{1}{4}f_{\mu \nu } \left[ {
\frac{{m_\gamma ^2 }}{\Delta }\left( {1 + \frac{{4\mu _s^2
}}{{\left( {\Delta  + 2m_H^2 } \right)}}} \right)} \right]f^{\mu \nu
}  + \frac{s}{2}\varepsilon ^{\mu \nu \lambda } b_\mu  \partial _\nu
b_\lambda. \label{ThreeCond230b}
\end{equation}

Once this is done, the above Hamiltonian constrained analysis can be
repeated step by step for this effective theory. Accordingly, the
potential for a pair of point-like opposite charges q located at
${\bf 0}$ and ${\bf L}$, in the  ${\raise0.5ex\hbox{$\scriptstyle
{\mu _S^2 }$} \kern-0.1em/\kern-0.15em
\lower0.25ex\hbox{$\scriptstyle {m_H^2 }$}} \to 0$ case, is given by
\begin{equation}
V = \frac{{q^2 }}{{2\pi }}K_0 (ML),  \label{ThreeCond235}
\end{equation}
where  $M^2  = {\raise0.7ex\hbox{${m_\gamma ^4 }$} \!\mathord{\left/
 {\vphantom {{m_\gamma ^4 } {s^2 }}}\right.\kern-\nulldelimiterspace}
\!\lower0.7ex\hbox{${s^2 }$}}$. We immediately see that,
unexpectedly, the confining potential between static charges
vanishes in this case.

\section{Final remarks}

To conclude, this work is a sequel to \cite{GaeteSpa09}, where we
have considered a three-dimensional extension of the recently
proposed Higgs-like model  \cite{Gabadadze4}, which describes a
condensed of charged scalars in a neutralizing background of
fermions. To do this, we have exploited a crucial point for
understanding the physical content of gauge theories, namely, the
correct identification of field degrees of freedom with observable
quantities. It was shown, that for the case of a term physical mass
$m_H^2 \phi ^ *  \phi$, a screening potential is recovered.
Interestingly enough, in the case of a "wrong-sign'' mass term $-
m_H^2 \phi ^ *  \phi$, unexpected features were found. It was
observed that the interaction energy is the sum of an
effective-Bessel and a linear potential, leading to the confinement
of static charges. However, when a Chern-Simons term is included,
the surprising result is that the theory describes an exactly
screening phase.

Acknowledgments\\
\\
One of us (PG) wants to thank the Field Theory Group of the CBPF for
hospitality and PCI/MCT for support. This work was supported in part
by Fondecyt (Chile) grant 1080260 (PG). (J.H-N) expresses his
gratitude to CNPq and the staff of the Department of Physics of the
Universidad T\'ecnica Federico Santa Mar\'{\i}a for the pleasant
stay.


\begin{thebibliography}{}

\bibitem{Nambu} Y. Nambu, Phys. Rev. {\bf D10}, 4262 (1974).

\bibitem{«t Hooft} G.'tHooft, {\it High Energy Physics}, ed. A. Zichichi
(Bologna: Editorice Compositori), (1975).

\bibitem{Mandelstam}  S. Mandelstam, Phys. Rep. {\bf C23} 245 (1976).

\bibitem{Gabadadze1} G. Gabadadze and R. A. Rosen, Phys. Lett. {\bf B666}, 277
(2008).

\bibitem{Gabadadze2} G. Gabadadze and R. A. Rosen, Phys. Lett. {\bf B658}, 266
(2008).

\bibitem{Gabadadze3} G. Gabadadze and R. A. Rosen, JCAP {\bf 0810}, 030 (2008).

\bibitem{Gabadadze4} G. Gabadadze and R. A. Rosen,  JCAP {\bf 0902}, 016 (2009).

\bibitem{GaeteSpa09} P. Gaete and E. Spallucci, Phys. Lett. {\bf B675}, 145 (2009).

\bibitem{GaeteWotzasek04}  P. Gaete and C. Wotzasek, Phys. Lett. {\bf B601},108 (2004).

\bibitem{GaeteGuen05} P. Gaete and E. I. Guendelman, Mod. Phys. Lett. {\bf A20}, 319 (2005).

\bibitem{GaeteSchmidt08} P. Gaete and I. Schmidt, arXiv: 0807.0816 [hep-th].

\bibitem{GaeteSpallucci08PRD} P. Gaete and E. Spallucci, Phys. Rev. {\bf D77}, 027702 (2008).

\bibitem{GaeteSpallucci08JPA} P. Gaete and E. Spallucci, J. Phys. {\bf A41}, 425401 (2008).

\bibitem{Appelquist} T. Appelquist and R. D. Pisarski, Phys. Rev. {\bf D23}, 2305 (1981).

\bibitem{Jackiw} R. Jackiw and S. Templeton, Phys. Rev. {\bf D23}, 2291 (1981).

\bibitem{Lee} K.-M. Lee, Phys. Rev. {\bf D49}, 4265 (1994).

\bibitem{Das} A. Das, \textit{Finite Temperature Field Theory, World Scientific}, Singapore, 1999.

\bibitem{Stone} M. Stone, \textit{The Quantum Hall Effect, World Scientific},
Singapore, 1992.

\bibitem{Helayel} J. A. Helay\"{e}l et al, Phys. Rev. {\bf D67}, 125011, (2003).

\bibitem{GaeteWot05} P. Gaete and C. Wotzasek, Phys. Lett. {\bf B625}, 365 (2005).

\bibitem{GaeteWot07} P. Gaete and C. Wotzasek, Phys. Rev. {\bf D75}, 057902 (2007).

\bibitem{GaeteHel08} P. Gaete and J. A. Helay\"{e}l, J. Phys. A: Math. Theor. {\bf 41}, 425401 (2008).

\bibitem{Polyakov} A. M. Polyakov, Nucl. Phys. {\bf B120}, 429 (1977).

\bibitem{Ezawa} Z. F. Ezawa, Phys. Lett. {\bf B85}, 87 (1987).

\bibitem{Orland} P. Orland, Nucl. Phys. {\bf B205}, 107 (1982).

\bibitem{Kondo} K.-I. Kondo, Phys. Rev. {\bf D58}, 085013 (1998).

\bibitem{Gaete99} P. Gaete, Phys. Rev. {\bf D 59}, 127702 (1999).

\bibitem{Pato} P. Gaete and I. Schmidt, Phys. Rev. {\bf D61}, 125002 (2000).

\bibitem{Pato2} P. Gaete, Z. Phys. C {\bf 76}, 355 (1997).

\bibitem{Dirac2} P. A. M. Dirac, \textit{The Principles of Quantum Mechanics},
4th ed. (Oxford University Press, Oxford, 1958); Can. J. Phys. {\bf
33}, 650 (1955).
\end{thebibliography}
\end{document}